\documentclass{article}

\PassOptionsToPackage{square, numbers}{natbib}
\usepackage[preprint]{neurips_2022}

\usepackage[utf8]{inputenc} 
\usepackage[T1]{fontenc}    
\usepackage{hyperref}       
\usepackage{url}            
\usepackage{booktabs}       
\usepackage{amsfonts}       
\usepackage{nicefrac}       
\usepackage{microtype}      
\usepackage{xcolor}         
\usepackage{graphicx}

\usepackage{amsmath}
\usepackage{amssymb}
\usepackage{amsthm}

\theoremstyle{definition}

\newcommand{\E}{\mathop{\mathbb{E}}}

\newcommand{\Var}{\operatorname{Var}}

\newcommand{\R}[0]{\mathbb{R}}

\newcommand*\diff{\mathop{}\!\mathrm{d}}

\usepackage{enumitem}
\usepackage{color}
\usepackage{listings}
\usepackage{algorithm}
\usepackage{algpseudocode}
\usepackage[edges]{forest}
\usepackage{pgfplots}
\usepackage{adjustbox}
\usepackage{xstring}
\usepackage{ifthen}
\usetikzlibrary{positioning}
\usepackage{stmaryrd}
\usepackage[toc, page]{appendix}

\makeatletter
\let\c@table\c@figure 
\let\ftype@table\ftype@figure 
\makeatother

\title{Multimodal Deep Reinforcement Learning for Portfolio Optimization}
\author{%
  Sumit Nawathe \quad Ravi Panguluri \quad James Zhang \quad Sashwat Venkatesh \\
  University of Maryland, College Park \\
  \texttt{\{snawathe, rpangulu, jzhang72, sashvenk\}@terpmail.umd.edu} \\
}

\pgfplotsset{compat=1.18}

\begin{document}
\maketitle

\begin{abstract}
  We propose a reinforcement learning (RL) framework that leverages multimodal data—including historical stock prices, sentiment analysis, and topic embeddings from news articles—to optimize trading strategies for S\&P100 stocks. Building upon recent advancements in financial reinforcement learning, we aim to enhance the state space representation by integrating financial sentiment data from SEC filings and news headlines and refining the reward function to better align with portfolio performance metrics. Our methodology includes deep reinforcement learning with state tensors comprising price data, sentiment scores, and news embeddings, processed through advanced feature extraction models like CNNs and RNNs. By benchmarking against traditional portfolio optimization techniques and advanced strategies, we demonstrate the efficacy of our approach in delivering superior portfolio performance. Empirical results showcase the potential of our agent to outperform standard benchmarks, especially when utilizing combined data sources under profit-based reward functions.
\end{abstract}


\newpage

\section{Introduction}
\label{sec:existing_literature}

Our group is seeking to develop a reinforcement learning agent to support portfolio 
management and optimization. Utilizing both empirical stock pricing data and alternative data such as SEC filings and news headlines, we create a more well-informed portfolio optimization tool. 

Our primary motivations for pursuing a reinforcement learning-based approach are as 
follows: firstly, reinforcement learning lends itself well to learning/opening in an online environment. The agent can interact with its environment, providing real-time feedback/responsiveness to allow for better results. Secondly, our approach involves incorporating alternative data to support the agent's decision-making process. Encoding this alt-data into the states matrix of the agent allows for the agent to make better decisions when it comes to adjusting portfolio weights. Finally, given that a reinforcement learning agent's decisions are modeled by a Markov Decision Process, we can easily provide different reward functions to account for a variety of investor preferences or restrictions. 

Our primary algorithmic technique is deep reinforcement learning, which uses deep neural networks to learn an optimal policy to interact with an environment and optimize performance towards a goal.
Formally, a reinforcement learning problem is an instance of a Markov 
Decision Process, which is a 4-tuple $(S, A, T, R)$: $S$ the state space 
(matrix of selected historical stock price and news data available to 
our model at a given time, $A$ the action space 
(portfolio weights produced by our model, under appropriate constraints), 
$T$ the transition function (how the state changes over time, modeled by our dataset), 
and $R$ (the reward function). The goal is to find a trading policy (function from $S \to A$) 
that maximizes future expected rewards. Most reinforcement learning research is 
spent on providing good information in $S$ to the model, defining a good reward 
function $R$, and deciding on a deep learning model training system to optimize rewards.

Much of the literature applying RL to portfolio optimization has arisen in the 
last few years. \cite{drl_mvo} 
use a look back at recent returns and a few market indicators 
(including 20-day volatility and the VIX), this paper implements a simple 
algorithm for portfolio weight selection to maximize the Differential Sharpe Ratio, 
a (local stepwise) reward function that approximates the (global) Sharpe Ratio of the 
final strategy. They compare their model with the standard mean-variance 
optimization across several metrics. \cite{drl_modern_portfolio_theory} 
applies reinforcement learning methods to 
tensors of technical indicators and covariance matrices between stocks. 
After tensor feature extraction using 3D convolutions and tensor decompositions, 
the DDPG method is used to train the neural network policy, and the algorithm 
is backtested and compared against related methods. \cite{rl_augmented_states} propose a method to augment the state space S of historical 
price data with embeddings of internal information and alternative data. 
For all assets at all times, the authors use an LSTM to predict the price movement,
which is integrated into S. When news article data is available, different NLP methods 
are used to embed the news; this embedding is fed into a HAN to predict price 
movement, which is also integrated into S for state augmentation. The paper applies 
the DPG policy training method and compares it against multiple baseline portfolios on 
multiple asset classes. It also addresses challenges due to environmental uncertainty, 
sparsity, and news correlations. \cite{drl_framework} rigorously discusses of how to properly incorporate 
transaction costs into an RL model. The authors also have a GitHub with implementations of their 
RL strategy compared with several others. \cite{learn_to_rank} explores news sentiment indicators including shock and trends and applies 
multiple learning-to-rank algorithms and constructs an automated trading system with strong performance. \cite{maps} takes advantage of reinforcement learning with multiple agents by defining a 
reward function to penalize correlations between agents, thereby producing multiple orthogonal 
high-performing portfolios.

\section{Data}
\label{ch:third} 
\label{stock_price_data_sourcing}

We outline the process of collecting, storing, and preprocessing all of the price data and alternative data used in our trading strategies.
The BUFN Computational Finance Minor program provided us access to Wharton Research Data Services (WRDS), specifically data from the Center for Research in Security Prices (CRSP).
Under the ideas and implementation of \cite{drl_framework}, we download basic stock price data -- close/high/low price, volume, and metadata --
for all stocks in the S\&P100 index from 2010 to 2020. We also download data for the S\&P500 value-weighted and equally-weighted indices for benchmark comparison \cite{crsp_data}.

All of the reinforcement learning agents we create will have access to historical company price data, as it broadly reflects the market's perceived value of a given company. However, we believe that using alternative sources will enhance our agents' decision-making process and provide value to our portfolio strategy. We aim to use two primary types of alternative data:
news headlines and SEC filings. We discuss our data sourcing, curation, and cleaning process at length in Section \ref{ch:third}. However, here we briefly motivate our usage of each in composing our multimodal dataset.

\subsection{SEC Filings Data}
\label{sec:getting_sec_data}

SEC filings include detailed information on a company's financial health and external risk factors directly from executives \cite{sec_how_to_read_10kq}.  SEC filings are filed under a single standard format by all publicly listed companies on a quarterly and yearly basis. Given the imposed structure of the documents and regular reporting periods, these filings provide a consistent source of external information. Further, we believe these filings could provide valuable future-looking insight into a company's operations that might not be directly immediately reflected in its stock price. The parts of SEC reports that we use are discussed in Section \ref{sec:getting_sec_data}. Section \ref{sec:sec_data_processing} discusses how to use the Loughran-McDonald sentiment dictionary to compute sentiment scores for each company on the date of filing release, our use of exponential decay when forward-filling these scores to future dates, and the quality of the data.

We used the EDGAR database \cite{sec_edgar} to download 10-K and 10-Q SEC filings for S\&P100 for the last 30 years.
The results are a set of HTML files taking up roughly 115GB of storage space, which we stored in Google Drive.
We built parsers to extract the key sections from both types of filings; in particular, Item 7/7A from the 10-K and Item 2 from the 10-Q.
This is the Management's Discussion and Analysis (MD\&A) section, which allows the company management to discuss
"the company's exposure to market risk, such as interest rate risk, foreign currency exchange risk, commodity price risk or equity price risk,", and "how it manages its market risk exposures" \cite{sec_how_to_read_10kq}.

\subsubsection{SEC Data Processing and Creating Tensors}
\label{sec:sec_data_processing}

To extract meaningful values from the text, we first parse and clean the SEC filing HTML documents so we can extract the raw text.
Then we use regular-expression-based text parsing to extract text from Item 1A and 7/7A, and Item 2 in 10-Qs.
We then construct a data frame, where each row contains the company ticker, the date of the filing, the extracted section name, and the text of the extracted section.
We attempted to replicate the FinBERT sentiment score procedure explained in Section \ref{finbert} for SEC filings.
However, issues were encountered both with the size of the dataset making applying FinBERT to these extracted sections too computationally intensive. There were also parsing issues due to the way the formatting irregularities in the filings.
Therefore, we use a modified process to create the sentiment tensors. We extract positive, negative, and neutral words as specified by the Loughran-McDonald sentiment dictionary, and then utilize the proportions similarly to the news embeddings using Equation \eqref{eq:value_embedding}.

The Loughran-McDonald sentiment dictionary is an academically maintained dictionary that lists business-specific words used to gauge the status of a firm. As documented in their 2011 paper in the \textit{Journal of Finance}, the dictionary contains a list of over 80,000 words, which each word flagged with a particular sentiment, such as "positive", "negative", "litigious", etc. We parse the SEC filings and tokenize them, then determine the proportion of positive, negative, and neutral words in the total filing, and then use \eqref{eq:value_embedding}, substituting in positive word proportion, negative word proportion, and neutral word proportion for positive sentiment probability, negative sentiment probability, and neutral sentiment probability, respectively. For this investigation, we utilize the 2023 version of the Master Sentiment Dictionary.

An issue we run into when incorporating SEC filing data is that they are recorded on an annual or
quarterly basis, which creates significant gaps between reporting dates. To help fill these, we again use exponential decay, defined in section \ref{eq:sentiment_decay}, and tune the $\gamma$ parameter during model training; once again, $\gamma \approx 0.8$ yielded good results.

\subsubsection{SEC Filings Dataset Statistics}
\label{sec:sec_filling_stats}
Our dataset contains data for 99 out of the 100 tickers in the S\&P 100, containing over 9,000 filings between 1994 and the present day, with the used subset consisting of roughly 6,100 filings.
Table \ref{table:sec_dist} shows some reported summary statistics on the distribution of SEC filings across the tickers:

\begin{table}[htbp]
    \centering
    \caption{Company SEC Filings Distribution}
    \begin{tabular}{l c}
        \toprule
        \textbf{Statistic} & \textbf{Value} \\
        \midrule
        Count & 99 \\
        Mean No. of Filings & 61.42 \\
        Standard Deviation & 10.18 \\
        Minimum Observations & 20 \\
        25th Percentile & 64 \\
        Median & 65 \\
        75th Percentile & 65 \\
        Maximum & 66 \\
        \bottomrule
    \end{tabular}
    \label{table:sec_dist}
\end{table}

Since there are only 4 filings per year, we use forward filling with decay to fill in the "missing" dates as in Equation \ref{eq:sentiment_decay}.
Given the addition and dropping of companies from the S\&P100, as well as some newer public companies joining, each company does not have the same number of filings over the time period.

\begin{figure}[H]
  \centering
  \includegraphics[width=.9\textwidth]{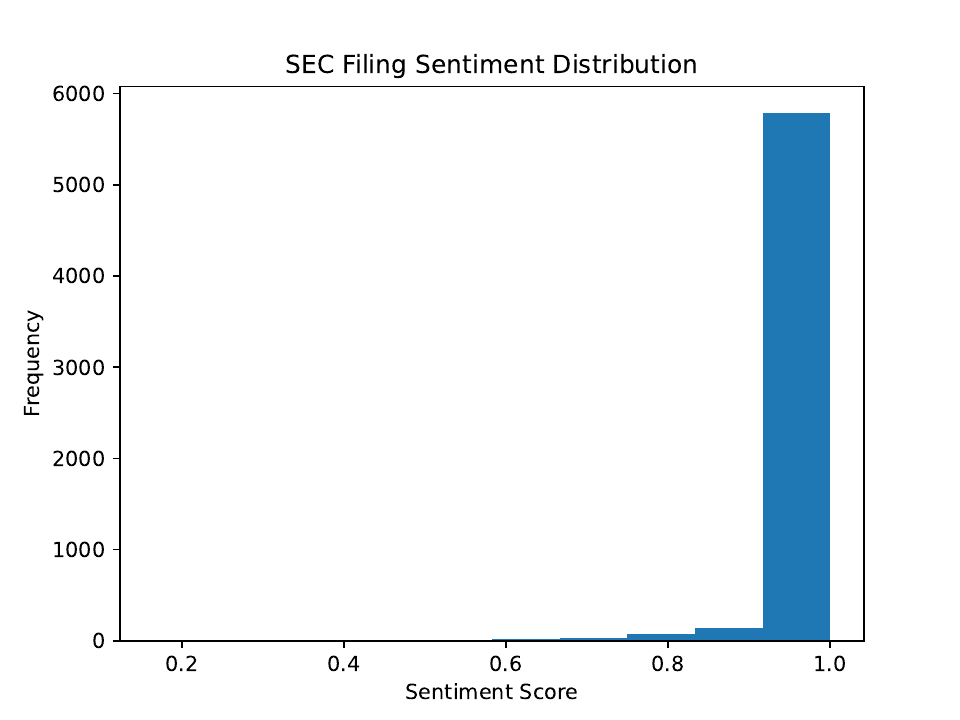}
  \caption{Frequency distribution of our novel SEC sentiment scores.}
  \label{fig:sec_dist}
\end{figure}

Figure \ref{fig:sec_dist} shows the distribution of sentiment scores.
There is a pronounced tail towards 1, indicating a strongly positive unimodal distribution, as compared to that of the news sentiment.
Since we utilize sections of the SEC filings that are written by the companies themselves,
companies likely aim to provide filings that suggest strong performance and future outlook.
In the dataset, we do observe some drops in sentiment, such as in times of financial crisis or bad market conditions, like in 2013 for some technology-based companies.

\subsection{News Headline Data}
\label{sec:stock_price_sourcing}

We incorporate company-specific news headlines in our agents' environment because they can reflect real-time shifts in investor perceptions that may take longer to be reflected in a company's price. Positive news such as acquisitions can drive stock prices up, while negative news such as leadership changes can have adverse effects. Therefore, having up-to-date sentiment information on each company in the trading universe could help our agent outperform its benchmarks. Acquisition of our news data is discussed in Section \ref{sec:stock_price_sourcing}. Section \ref{sec:news_processing} discussed how we obtain FinBERT sentiment scores, our novel function for creating sentiment embeddings, our process for forward-filling sentiment data using exponential decay, and the quality of the new data.


\subsubsection{Daily Financial Headlines Dataset}
The dataset we use in this project is Daily Financial News for 6000+ Stocks which was downloaded via Kaggle \cite{financial_news}.
This dataset contains scraped headline data for over 6000 stocks listed on the NYSE exchange from 2009 to 2020. 
There are two main files within this dataset that we use. The first is \texttt{raw\_analyst\_ratings.csv}, which only contains scraped data from a prominent financial news publisher Benzinga.
The other file \texttt{raw\_partner\_headlines.csv} contains scraped headline data from other smaller publishers that partner with Benzinga. Each row of the datasets contains a headline, the base article URL, the publisher, the date and time of publication, and the stock ticker symbol. We concatenate the headline data from each file to create a single unified dataset that contains all available news headlines in our trading period for all S\&P 100 stocks.

\subsubsection{News Data Processing and Creating Tensors}
\label{sec:news_processing}
\label{finbert}
\label{newstensors}

Over the full trading period (2010-2020), headlines for S$\&P$ 100 companies are fed into pre-trained FinBERT. 
The model then generates probabilities of the content having a positive, negative, or neutral sentiment. For news headlines, we developed a novel function to extract a single embedding for a stock on a given day. 

The function that we created is:
\begin{equation}\label{eq:value_embedding}
  \texttt{Value}_{\texttt{Embedding}} = \tanh\Biggl( \frac{\frac{\texttt{positive sentiment probability}}{\texttt{negative sentiment probability}}}{\texttt{neutral sentiment probability}} \Biggr)
\end{equation}
This approach captures the sentiment polarity by measuring the ratio between positive and negative sentiment in the numerator term. Dividing this ratio by the neutral sentiment probability imposes a penalty in the case that a headline is likely neutral. In that case, even if the ratio between negative and positive sentiment probabilities is high, we lose the information that the sentiment of the headline is likely neutral. Finally, our approach uses the tanh for normalization changing the domain of sentiment scores to be between -1 and 1. A sentiment score close to 1 can be interpreted as a positive sentiment, a score close to 0 can be interpreted as neutral, and a score close to -1 can be interpreted as negative.

An issue we run into with news data is irregular reporting dates and significant gaps in data reporting, which is described in more detail in section \ref{newsstats}.
To address some gaps in news data reporting, we apply exponential decay to the sentiment scores on report dates. Formally,
\begin{equation}\label{eq:sentiment_decay}
  y = a(1 - \gamma)^t
\end{equation}
where $a$ represents the company's sentiment score on the most recent reporting date,
$t$ represents the time (in days) between the last report date and the current day,
and $\gamma$ is a constant between 0 and 1 representing the daily decay factor. In our training process, we tune $\gamma$ as a hyperparameter to see what rate of decay yields the best-performing agents; we found that $\gamma \approx 0.8$ worked well for us.

From the concatenated dataset of news headline data from each publisher, as described in the "News Data" section, we feed the dataset (loaded into a Pandas DataFrame) through a multi-stage pipeline. 
The first step is to scrape the current S\&P 100 companies and then filter the dataset down to only include headlines from companies in the S\&P 100. 
We introduce a custom dataset class called "NewsHeadlines," implemented in the PyTorch framework, designed for efficiently handling news headline data. 
The class takes a dataset and a user-defined tokenizer which will pre-process headlines in batches to be fed into FinBERT. 
In the class, we implement an iterator function \texttt{\_getitem}, which takes the raw headline data as input and returns an encoding for the batch of headlines after tokenization. 
Then given the large size of the dataset, we create a "Dataloader" object, implemented in PyTorch, which feeds our dataset into the model in small batches. 

To obtain the output tensors corresponding to the sentiment probabilities, we iterate over the batches, applying FinBERT to classify each headline and from the raw logits using the softmax activation function to a vector of probabilities.
Then for each batch, we save the tensors into separate files.

\subsubsection{News Dataset Statistics}
\label{newsstats}

Our dataset contains data for 84 out of the total 100 tickers in the S\&P 100,
and it contains 70,872 entries containing the sentiment embedding of news for a company on a given day.
Table \ref{table:news_dist} displays some summary statistics on the distribution of news reports across the tickers.

\begin{table}[htbp]
    \centering
    \caption{Company News Reporting Date Distribution}
    \begin{tabular}{l c}
        \toprule
        \textbf{Statistic} & \textbf{Value} \\
        \midrule
        Count & 84 \\
        Mean No. of Reporting Dates & 843.714 \\
        Standard Deviation & 508.209 \\
        Minimum Observations & 1 \\
        25th Percentile & 393.250 \\
        Median & 905.000 \\
        75th Percentile & 1198.500 \\
        Maximum & 1829 \\
        \bottomrule
    \end{tabular}
    \label{table:news_dist}
\end{table}

Note that our median ticker only has news reports on 905 of the total trading dates and since there are 16 tickers for which we have no sentiment data, our dataset is still sub-optimal for developing an agent. Our forward-filling process does address some of the gaps in our data, however, our coverage is still incomplete.
This is an important consideration when examining the results of our work.

\begin{figure}[H]
  \centering
  \includegraphics[width=.9\textwidth]{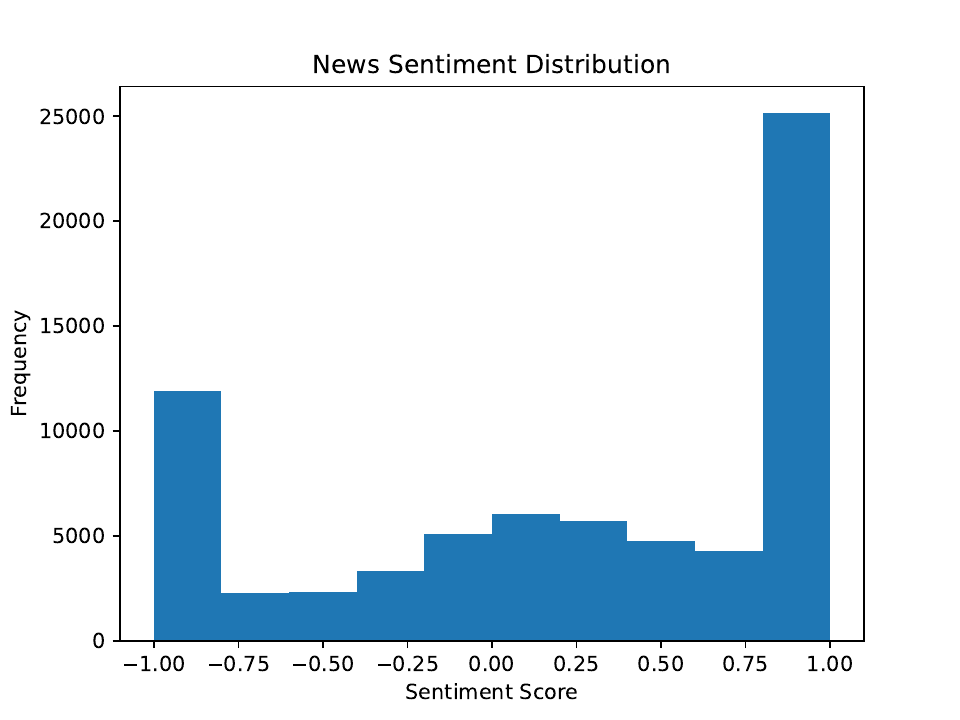}
  \caption{Frequency distribution of our novel news sentiment scores.}
  \label{fig:news_dist}
\end{figure}

Figure \ref{fig:news_dist} shows the distribution of sentiment scores across the articles.
News sentiment has a bimodal distribution: much of the headlines are interpreted as either negative or positive, but news headlines are relatively neutral, closer to 0, or more evenly distributed. 
This indicates that the headlines display strong enough sentiment that they could inform and change the actions of our reinforcement learning agents.

\section{Methodology}

We will be implementing and improving on the methodologies of several of the above papers. 
We develop a reinforcement learning system that utilizes multiple periods to achieve strong out-of-sample trading performance. Our final architecture is most similar to papers \cite{rl_augmented_states} and \cite{drl_framework}.

\subsection{Markov Decision Process Problem Formulation}

Paper \cite{rl_augmented_states} includes the following diagram, which is very close to our desired architecture:

\begin{center}
\includegraphics[width=13cm]{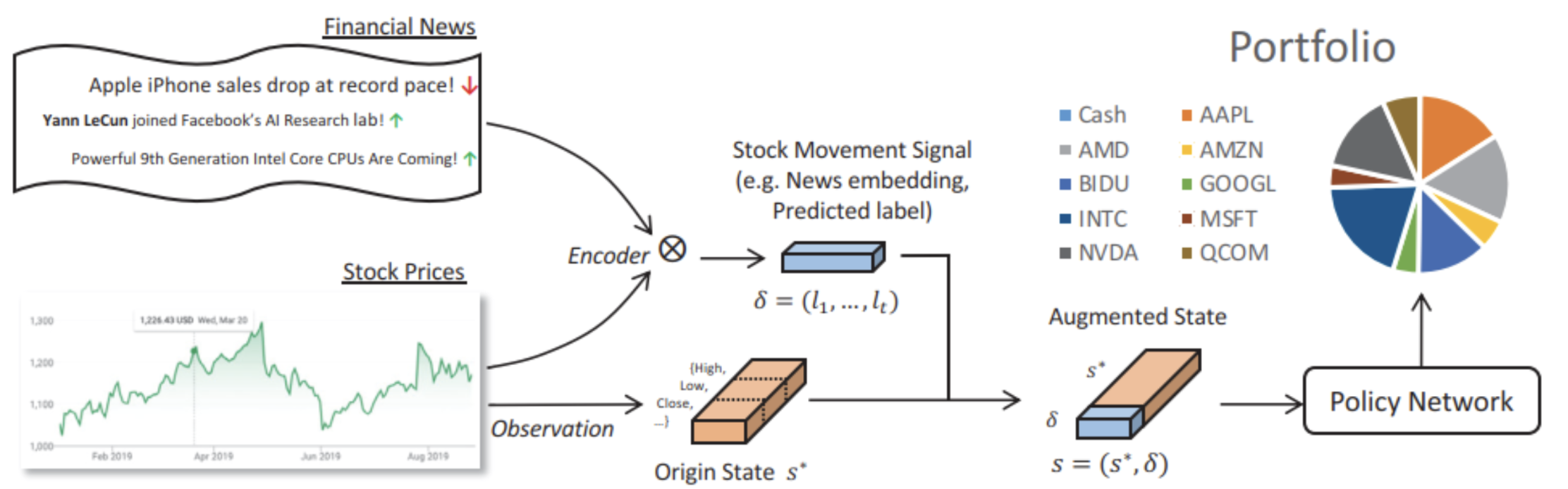}
\end{center}

An explanation of this diagram: at time t, the origin state $S^*$ is a 3D tensor of dimensions $U \times H \times C$ which contains historical price data. $U$ is the size of our universe (for example, for the S$\&$P100, $U = 100$). 
$H$ is the size of history we are providing (if we are providing 30-day history, then $H = 30$). 
$C$ is a categorical value representing the close/high/low price. This format of $S^*$ allows us to store, 
for example, the last 30 days of stock price data for all companies in the S$\&$P100, for any given day. 
In addition to this, we have news information $\delta$, obtained from financial news headlines for that day, 
processed through a pre-trained encoder. This information is added as an additional channel to $S^*$ to create the full state tensor $S = (S^*, \delta)$.

In our architecture, the signal information $\delta$ will be composed of process SEC and sentiment news indicators. The structure of the state $S$ will remain a 3D tensor, as further described in Sections \ref{sec:eiie_policies}. Each row of $S$ will represent a different stock in our universe, and along that row will be all of the price and alternative data for the past several days. (As discussed in the literature, the straightforward concatenation of price data and news embeddings does not affect the ability of the neural network-based agent to learn.)

Regarding the reward function $R$, we plan to experiment with both the profit reward function used in paper \cite{rl_augmented_states}, as well as the Differential Sharpe Ratio developed in paper \cite{drl_mvo}.
The former is simply the change in the portfolio value over the last period based on the weights (action) provided by the agent; the latter attempts to make the cumulative reward approximate the Sharpe ratio over the entire period.

In all of the papers, the action space $A$ is a length $m+1$ vector such that the sum of all elements is $1$,
where there are $m$ stocks in our universe (the other weight is for the risk-free asset).
Each action represents the agent's desired portfolio for the next time period and is computed based on the
state at the end of the previous period. We will experiment with short-selling and leverage restrictions
(which put lower and upper bounds on the weight components, respectively).

In summary, our project aims to implement and replicate the approach used in \cite{rl_augmented_states}, with some modifications to $S$ and $R$ as previously described. We will conduct experiments on alternative data 
sources, feature extraction methods, and reward functions (both custom and from other papers listed) to find a good combination that allows this approach to work well on S$\&$P100 stocks; this comprises our novel extension/contribution.



\subsection{Strategy Benchmarking}

Our final model architecture is compared against several benchmark financial portfolio selection 
models. Among these will be a naive equally weighted portfolio, a naive buy-and-hold portfolio, and holding the asset with the best historical Sharpe Ratio. These simple strategies are defined in Section \ref{sec:benchmark_section}. In addition, we test our agents against two more advanced benchmark strategies: OLMAR and WMAMR, which are defined in \ref{AppendixB}

We will compare our returns 
in-sample and out-of-sample plots, as well as our relative performance on portfolio statistics 
including cumulative return, Sharpe Ratio, Sortino Ratio, drawdown, etc. The experiment sections of the papers we discuss in Section \ref{sec:existing_literature} provide a strong reference for our methodological comparison.

\subsection{Specialization to Our Application}

In the portfolio optimization setting, our RL agent seeks to produce an optimal set of portfolio weights given all the information it knows.
Assume that there are $m$ tradable stocks in our universe and one risk-free asset.
The action space $A = \left\{a \in \R^{m+1} \:|\: \sum_i a_i = 1\right\}$ is the set of all possible portfolio weights.

The state space $S$ encompasses all information available to the agent when it is asked to make a portfolio allocation decision at a given time.
Depending on what information is provided, this could include past performance of the strategy, historical stock prices, encoded news information for
select/all tickers in the universe, or some combination of these. For most of the scenarios we consider, $S \in \R^{n \times N}$ is a matrix, where
each row corresponds to a different stock ticker, and along that row, we find the past few weeks of historical price data as well as some
aggregate of news sentiment indicators/scores.

The transition function $T$ is a delta function since state transitions are deterministic.
The environment uses the weights provided by the agent to reallocate the portfolio, computes the new portfolio value,
and reads in new historical stock data and news data points to form the next state (for the next time period) which is provided to the agent.
(The exact form of $T$ is not needed; it is implicitly defined by the deterministic environment updates.)

The reward function $R$ should be such that it encourages the agent to produce good portfolio weights.
One simple reward function is pure profit: $R(s_t, a_t)$ is how much profit is gained to portfolio allocation $a_t$ during time interval $[t, t+1)$.
Another possible reward function is the Differential Sharpe ratio (as described in section \ref{diff_sharpe_ratio_section}), which urges the agent to make portfolio allocations to maximize its total Sharpe ratio.

\subsection{Differential Sharpe Ratio}
\label{diff_sharpe_ratio_section}

\cite{drl_mvo} utilizes the Differential Sharpe Ratio to implement and evaluate a reinforcement learning agent.
The Differential Sharpe Ratio is based on Portfolio Management Theory, and is developed in the author's previous works \cite{diff_sharpe_ratio_paper} and \cite{diff_sharpe_ratio_book}.
We briefly review the theory developed in both sources.

The traditional definition of the Sharpe Ratio is the ratio of expected excess returns to volatility.
If $R_t$ is the return of the portfolio at time $t$, and $r_f$ is the risk-free rate then
\begin{align*}
  \mathcal{S} = \frac{\mathbb{E}_t[R_t] - r_f}{\sqrt{\Var_t[R_t]}}
\end{align*}

The primary application of the Sharpe Ratio is to analyze strategy once all data is collected.
Furthermore, modern portfolio theory aims to maximize the Sharpe Ratio over the given period, or equivalently, to maximize the mean-variance utility function).

Unfortunately, this will not work for a reinforcement learning agent. The agent must be given a reward after every time step,
but the traditional Sharpe ratio is only calculated at the end.

The Differential Sharpe Ratio attempts to remedy this by approximating a change in the total Sharpe ratio up to that point.
By summing together many of these incremental changes (though approximate), the cumulative rewards are an approximation of the total
Sharpe ratio over the complete period.

The approximation works by updating moment-based estimators of the expectation and variance in the Sharpe Ratio formula.
Let $A_t$ and $B_t$ be estimates of the first and second moments of the return $R_t$ up to time $t$.
After time step $t$, having obtained $R_t$, we perform the following updates:
\begin{align*}
  \Delta A_t = R_t - A_{t-1} && A_t = A_{t-1} + \eta \Delta A_t \\
  \Delta B_t = R_t - B_{t-1} && B_t = B_{t-1} + \eta \Delta B_t
\end{align*}
where $A_0 = B_0 = 0$ and $\eta \sim 1/T$ is an update parameter, where there are $T$ total time periods.
These updates are essentially exponential moving averages.

Let $S_t$ be an approximation of the Sharpe Ratio up to time $t$ based on estimates $A$ and $B$. That is,
\begin{align*}
  S_t = \frac{A_t}{\sqrt{B_t - A_t^2}}
\end{align*}
The definition here ignores the risk-free rate term. $K_\eta$ is a normalization constant to ensure an unbiased estimator.

Pretend that at the update for time $t$, $A_{t-1}$ and $B_{t-1}$ are constants,
and $R_t$ is also a known constant. Then the updates to $A_t$ and $B_t$ only depend on the time step parameter $\eta$.
Indeed, if $\eta = 0$, then $A_t = A_{t-1}$ and $B_t = B_{t-1}$, so $S_t = S_{t-1}$.
Now consider varying $\eta$; expanding the Sharpe ratio estimator formula in Taylor series gives
\begin{align*}
  S_t \approx S_{t-1} + \eta \frac{\diff S_t}{\diff \eta}\Big|_{\eta = 0} + o(\eta^2)
\end{align*}

If $\eta$ is small, the final term is negligible, so this formula gives us an exponential-moving-average update for $S_t$.
The Differential Sharpe Ratio is defined to be a proportional derivative in that expression. With some tedious calculus, we find that
\begin{align*}
  D_t &= \frac{\diff S_t}{\diff \eta} = \frac{\diff}{\diff\eta}\left[ \frac{A_t}{\sqrt{B_t - A_t^2}} \right] = \frac{\frac{\diff A_t}{\diff \eta} \sqrt{B_t - A_t^2} - A_t \frac{\frac{\diff B_t}{\diff \eta} - 2 A_t \frac{\diff A_t}{\diff \eta}}{2 \sqrt{B_t - A_t^2}}}{B_t - A_t^2} \\
  &= \frac{\Delta A_t \sqrt{B_t - A_t^2} - A_t \frac{\Delta B_t - 2 A_t \Delta A_t}{2 \sqrt{B_t - A_t^2}}}{B_t - A_t^2}
  = \frac{B_t \Delta A_t - \frac{1}{2} A_t \Delta B_t}{(B_t - A_t^2)^{3/2}}
\end{align*}

This reward function is simple to implement in an environment. The authors of the original papers provide
experimental support for the value of this reward function in a reinforcement learning setting.

\subsection{Transaction Costs}
\label{transaction_costs_section}

\cite{drl_framework} contains an excellent walkthrough of the mathematics for modeling transaction costs in RL environment updates. We provide a shortened version here.

Suppose we have $m$ tradable assets and a risk-free asset. Let $\mathbf v_t = (1, v_{1,t}, v_{2,t}, \ldots v_{m,t}) \in \R^{m+1}$ be the
prices of the assets at time $t$ (the first entry is the risk-free asset). The raw return vector is defined as
$\mathbf y_t = \mathbf v_t \varoslash \mathbf v_{t-1} \in \R^{m+1}$, where division is element-wise. Suppose the portfolio
weight vector during period $t$ is $\mathbf w_t \in \R^{m+1}$, and let the value of the portfolio value at time $t$ be $p_t$.
If we were not considering transaction costs, then the portfolio return would be $\frac{p_t}{p_{t-1}} = \mathbf y_t \cdot \mathbf w_t$.

Unfortunately, buying and selling assets incur transaction costs. Let $\mathbf w_t' \in \R^{m+1}$ be the effective portfolio weights
at the end of time $t$ (it has changed from $\mathbf w_t$ due to the price changes). We have
\begin{align*}
  \mathbf w_t' = \frac{\mathbf y_t \odot \mathbf w_t}{\mathbf y_t \cdot \mathbf w_t}
\end{align*}
where $\odot$ is element-wise multiplication. Between time $t-1$ and time $t$, the portfolio value is also
adjusted from $p_{t-1} \in \R$ to $p_t' = p_{t-1} \mathbf y_t \cdot \mathbf w_{t-1}$. Let $p_t$ be the portfolio's value
after transaction costs, and let $\mu_t \in \R$ be the transaction cost factor, such that $p_t = \mu_t p_t'$.
We can keep track of the relevant time of each variable with the following diagram:

\begin{center}
  \includegraphics[width=13cm]{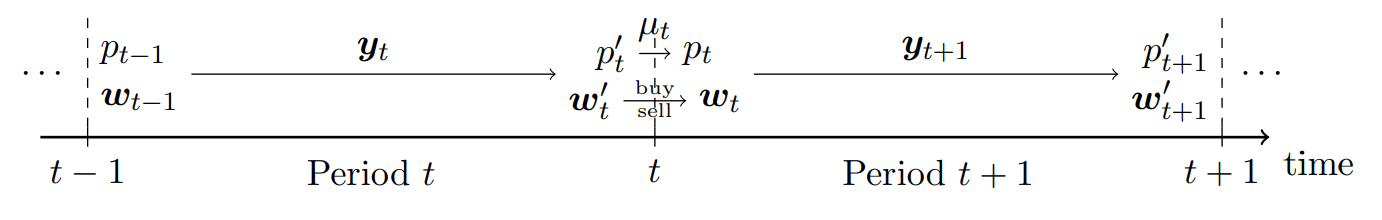}
\end{center}

In this paradigm, the final portfolio value at time $T$ is
\begin{align*}
  p_T = p_0 \prod_{t=1}^T \frac{p_t}{p_{t-1}} = p_0 \prod_{t=1}^T \mu_t \mathbf y_t \cdot \mathbf w_{t-1}
\end{align*}
The main difficulty is in determining the factor $\mu_t$ since it is an aggregate of all the transaction cost penalties.

Let $c_s \in [0, 1)$ be the commission rate for selling. We need to sell some amount of asset $i$ if 
there is more of asset $i$ in $\mathbf w_t'$ than in $\mathbf w_t$ by dollar value. Mathematically, this condition
is $p_t' w_{i,t}' > p_t w_{t,i}$, which is equivalent to $w_{i,t}' > \mu_t w_{i,t}$. Thus, the total amount of
money raised from selling assets is
\begin{align*}
  (1-c_s) p_t' \sum_{i=1}^m (w_{i,t}' - \mu_t w_{i,t})^+
\end{align*}
where $(\cdot)^+ = \max\{0, \cdot\} = \mathrm{ReLU}(\cdot)$. This money, as well as the money from adjusting the cash reserve
from $p_t' w_{0,t}'$ to $p_t w_{0,t}$, is used to purchase assets according to the opposite condition.
Let $c_p \in [0, 1)$ be the commission rate for purchasing. Equating the amount of money available from selling/cash and the amount of money
used for purchasing assets yields
\begin{align*}
  (1-c_p)\left[ w_{0,t}' - \mu_t w_{0,t} + (1-c_s) p_t' \sum_{i=1}^m (w_{i,t}' - \mu_t w_{i,t})^+ \right] = p_t'\sum_{i=1}^m (\mu_t w_{i,t} - w_{i,t}')+
\end{align*}
Moving terms around and simplifying the ReLU expressions, we find that $\mu_t$ is a fixed point of the function $f$ defined as:
\begin{align*}
  \mu_t = f(\mu_t) = \frac{1}{1 - c_p w_{0,t}}\left[ 1 - c_p w_{0,t}' - (c_s + c_p - c_s c_p) \sum_{i=1}^m (w_{i,t}' - \mu_t w_{i,t})^+ \right]
\end{align*}
The function $f$ is nonlinear. However, for reasonable values of $c_s$ and $c_p$, $f$ is both a monotone increase and a contraction, so iteratively computing values of $f$ can find its unique fixed point. This procedure is fairly efficient and easy to implement.

\subsection{EIIE Policies}
\label{sec:eiie_policies}
The second main contribution of \cite{drl_framework} is to create the framework of
Ensemble of Identical Independent Evaluators (EIIE) for a policy.
The principle is to have a single evaluation function that, given the price history and other data for a single asset,
produces scores representing potential growth for the immediate future. This same function is
applied to all assets independently, and the Softmax of the resulting scores becomes the portfolio's weights.

Written mathematically: let $X_{i,t}$ be the historical price information for asset $i$ at time $t$, and
let $w_{i,t}$ represent the portfolio weights for asset $i$ at time $t$.
Let $\alpha$, $\beta$, and $\gamma$ be trainable parameters.
First, we define a function $f_\alpha$ to extract features from each asset's price data $X_{\times, t}$. This function is applied to each asset individually.
The agent also needs to incorporate the previous portfolio weights into its action, to deal with the transaction costs as in section \ref{transaction_costs_section}.
We define a second function $g_\beta$ that takes the features produced by $f_{\theta_1}$, as well as the previous portfolio weights, to produce new weights.
Then the portfolio weights for the next period are
\begin{align*}
  w_{t+1} = \mathop{\mathrm{Softmax}}(g_\beta(f_\alpha(X_{1,t}), w_{1,t}), \ldots, g_\beta(f_\alpha(X_{m,t}), w_{m,t}), g_\beta(\gamma, w_{m+1,t}))
\end{align*}

(Note that there are $m$ tradable assets in our universe and that $w_{m+1, t}$ is the weight for the risk-free asset).

The form of $g_\beta$ is fairly arbitrary; the authors take it to be an MLP neural network.
The form of $f_\alpha$ is more interesting. The authors of \cite{drl_framework} suggest two forms: a CNN and an RNN/LSTM.

For the CNN they provide the following diagram of the EIIE system:

\begin{center}
  \includegraphics[width=0.9\textwidth]{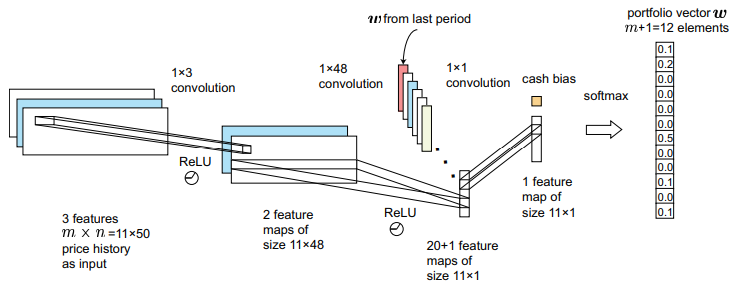}
\end{center}

For the RNN they provide a similar diagram:

\begin{center}
  \includegraphics[width=0.9\textwidth]{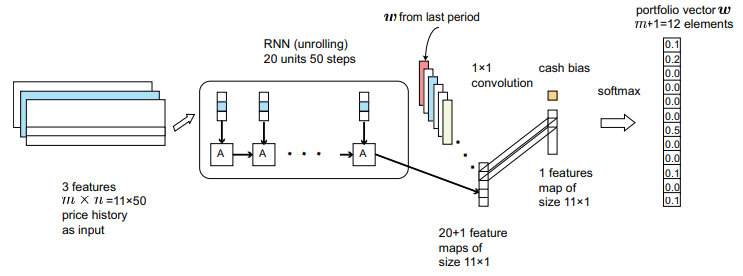}
\end{center}

The three channels in the state matrix on the left are the high, low, and closing prices for each asset on each day in the historical sliding window.

The authors claim that this framework beats all of their benchmark strategies in the cryptocurrency market. The architecture remains essentially the same in our implementation. Due to the use of multiple channels in the state tensor, it makes it easy to
add additional channels for alternative data sources, such as news sentiment.

\section{Empirical Results}

We run several experiments testing various combinations of alternative data usage, reward function, and policy type. Our training period is from the start of 2010 to the end of 2017, and the test period is from the start of 2018 to the end of 2019. These 10 years were chosen as it is the largest intersection of all data sources available to us. All of the strategies described in this chapter were trained and tested during these periods; only test plots and statistics are shown. For testing, we start each trained strategy with \$1 and let it run over the test period, and plot the value of the strategy's portfolio over time. Transaction costs are held at 1\% throughout.

The cumulative returns plots for each group of strategies for this section have been collected in \ref{chap:gallery}. We choose to only display comparative summary tables in this section because they are easier to interpret.

\subsection{Benchmarks Portfolios}
\label{sec:benchmark_section}
We begin by reviewing the benchmarks highlighted in the overview:
\begin{itemize}
  \item Naive Equal: A simple strategy that trades to maintain equal weights across all assets (and the risk-free asset).
  \item Equal Buy-and-Hold: It initially buys equal amounts of all assets, and then does no further trading.
  \item Best Historical Sharpe: Puts all money in the single asset that had the best Sharpe ratio during the training period.
  \item OLMAR and WMAMR: Described in Appendix A. 
  \item S\&P500: The returns of the index, as provided by CRSP (see Section \ref{stock_price_data_sourcing}).
\end{itemize}

\begin{figure}[!htb]
    \centering
    \includegraphics[width=0.8\textwidth]{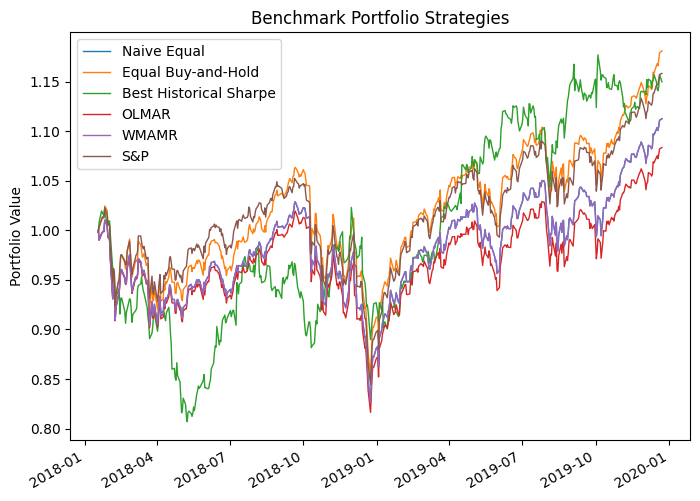}
    \caption{Performance of benchmark strategies on test period}
    \label{fig:benchmark_results}
\end{figure}

The performance of the benchmarks is shown in Table \ref{tab:benchmark_table} and Figure \ref{fig:benchmark_results}.

\begin{table}[htbp]
    \centering
    \caption{Benchmark Portfolio Strategies}
      \begin{tabular}{lcccc}
      \toprule
            & Net Profit & Sharpe Ratio & Sortino Ratio & Max Drawdown \\
      \midrule
      Naive Equal & 0.112583 & 0.392352 & 0.473881 & 0.197294 \\
      Equal Buy-and-Hold & 0.181138 & 0.595225 & 0.716195 & 0.198039 \\
      Best Historical Sharpe & 0.149857 & 0.475577 & 0.683296 & 0.210250 \\
      OLMAR & 0.083634 & 0.300906 & 0.364244 & 0.199240 \\
      WMAMR & 0.112565 & 0.392310 & 0.473818 & 0.197294 \\
      S\&P & 0.158347 & 0.552242 & 0.675382 & 0.197728 \\
      \bottomrule
      \end{tabular}%
    \label{tab:benchmark_table}%
  \end{table}%

Most benchmarks underperform the S$\&$P index over the trading period in terms of profitability.
The simple Equal Buy-and-Hold is the best performer by net profit, Sharpe Ratio, and Sortino Ratio. Both of the trading strategies OLMAR and WMAMR also lag behind the index. The Best Historical Sharpe stock also underperforms the market and has much higher volatility indicated by its significantly lower Sharpe Ratio compared with the S\&P with similar profitability. All strategies have a similar max drawdown due to the market decline in late 2018 / early 2019.

\subsection{RL Portfolios: Historical Price Data}
\label{sec:hist_price_results}
Our first set of strategies resembles the implementation of \cite{drl_framework} as described in Section \ref{sec:eiie_policies}. At each time, we provide the agent with a tensor of close/high/low stock prices for the past few weeks. We test the policies of CNN EIIEs, RNN EIIE, and a standard MLP neural network on the entire tensor. We also test using both the Differential Sharpe ratio reward and the profit reward.

\begin{figure}[!htb]
    \centering
    \includegraphics[width=0.8\textwidth]{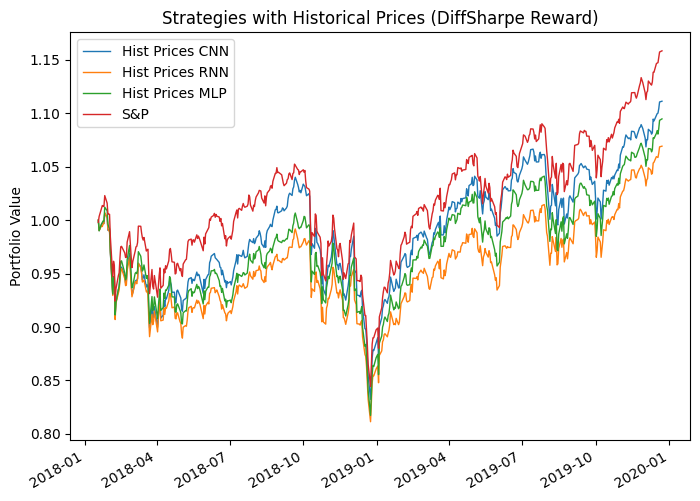}
    \caption{Performance of strategies using historical price data with Differential Sharpe reward}
    \label{fig:historical_diffsharpe_results}
\end{figure}

\begin{figure}[!htb]
    \centering
    \includegraphics[width=0.8\textwidth]{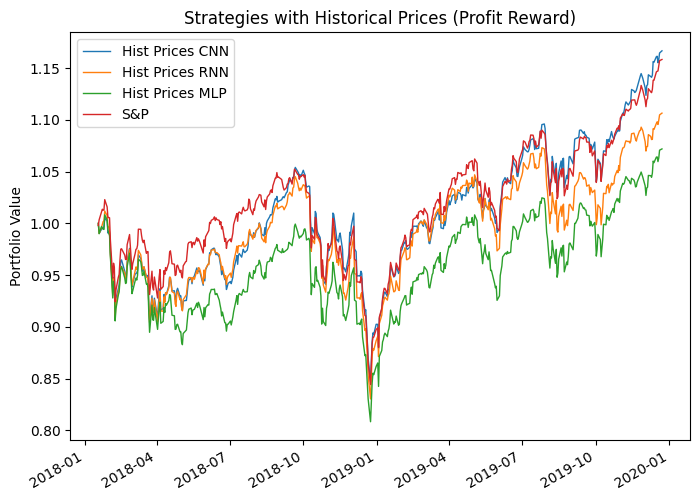}
    \caption{Performance of strategies using historical price data with Profit reward}
    \label{fig:historical_profit_results}
\end{figure}

The results with the Differential Sharpe Ratio reward can be found in Table \ref{tab:hist_diff_table} and Figure \ref{fig:historical_diffsharpe_results}, while results with the Profit reward are found in Table \ref{tab:hist_profit_table} and Figure \ref{fig:historical_profit_results}.

\begin{table}[htbp]
    \centering
    \caption{Performance of Strategies with Historical Prices (Differential Sharpe Reward)}
    \begin{tabular}{lcccc}
        \toprule
        & Net Profit & Sharpe Ratio & Sortino Ratio & Max Drawdown \\
        \midrule
        Hist Prices CNN & 0.111205 & 0.396336 & 0.477160 & 0.199721 \\
        Hist Prices RNN & 0.069154 & 0.262906 & 0.318161 & 0.194177 \\
        Hist Prices MLP & 0.094686 & 0.338142 & 0.409170 & 0.188082 \\
        S\&P   & 0.158347 & 0.552242 & 0.675382 & 0.197728 \\
        \bottomrule
    \end{tabular}%
    \label{tab:hist_diff_table}%
\end{table}%

\begin{table}[htbp]
    \centering
    \caption{Performance of Strategies with Historical Prices (Profit Reward).}
    \begin{tabular}{lcccc}
        \toprule
        & Net Profit & Sharpe Ratio & Sortino Ratio & Max Drawdown \\
        \midrule
        Hist Prices CNN & 0.166717 & 0.561471 & 0.676669 & 0.196557 \\
        Hist Prices RNN & 0.106518 & 0.369764 & 0.454653 & 0.205465 \\
        Hist Prices MLP & 0.071937 & 0.261355 & 0.322155 & 0.198329 \\
        S\&P   & 0.158347 & 0.552242 & 	0.675382 & 0.197728 \\
        \bottomrule
    \end{tabular}%
    \label{tab:hist_profit_table}%
\end{table}%

From Table \ref{tab:hist_diff_table}, strategies using the Differential Sharpe ratio reward all have mediocre returns, significantly underperforming the S\&P index across all statistics. This indicates that the policy gradient optimization process is not as effective in optimizing the policy. Indeed, the Differential Sharpe Ratio is a difficult reward for the policy to learn given the agent's limited information. We note that the CNN appears to perform better than both the RNN and MLP policies here.

In contrast, in Table \ref{tab:hist_profit_table}, the CNN and RNN EIIE policies show reasonable performance, but the MLP does not. The CNN slightly outperforms the S\&P index, while the MLP policy deteriorates significantly. Profit reward is relatively easy to learn, and so the small (and therefore more noise-robust) EIIE models fare well, while the MLP model overfits. and does not see improvements.

The trends seen by statistical analysis, and also displayed visually in Figures \ref{fig:historical_diffsharpe_results} and \ref{fig:historical_profit_results}, are common themes throughout the rest of this analysis.

\subsection{RL Portfolios: Price + SEC Data}
We now keep the same types of policies and reward functions but augment the state tensor with additional channels for the SEC sentiment scores for every asset on every day in the lookback window.

\begin{figure}[!htb]
    \centering
    \includegraphics[width=0.8\textwidth]{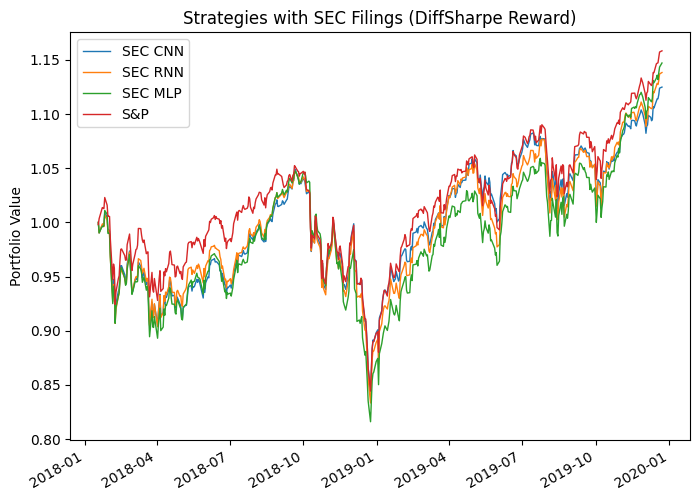}
    \caption{Performance of strategies using price and SEC data with Differential Sharpe reward}
    \label{fig:sec_diffsharpe_results}
\end{figure}

\begin{figure}[!htb]
    \centering
    \includegraphics[width=0.8\textwidth]{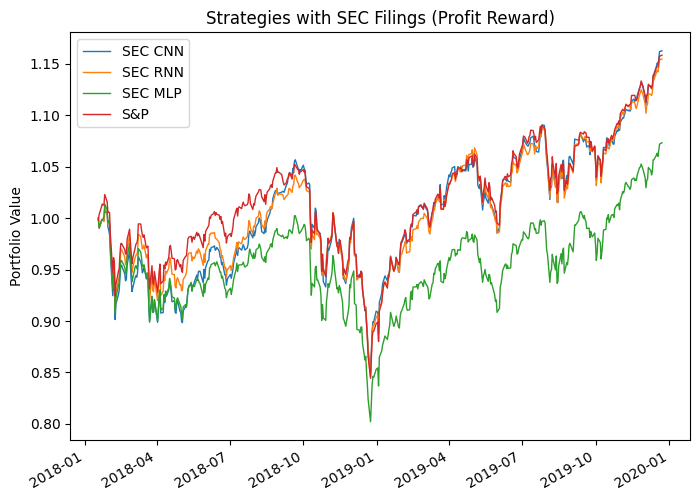}
    \caption{Performance of strategies using price and SEC data with Profit reward}
    \label{fig:sec_profit_results}
\end{figure}

The results with the Differential Sharpe Ratio reward can be found in Table \ref{tab:sec_diffsharpe_results} and Figure \ref{fig:sec_diffsharpe_results}, while results with the Profit reward are found in Table \ref{tab:sec_profit_results} and Figure \ref{fig:sec_profit_results}.

\begin{table}[htbp]
    \centering
    \caption{Strategies with SEC Filings (Differential Sharpe Reward)}
      \begin{tabular}{lcccc}
      \toprule
            & Net Profit & Sharpe Ratio & Sortino Ratio & Max Drawdown \\
      \midrule
      SEC CNN & 0.124870 & 0.439863 & 0.529951 & 0.195347 \\
      SEC RNN & 0.138309 & 0.464039 & 0.568351 & 0.204159 \\
      SEC MLP & 0.147103 & 0.482788 & 0.581065 & 0.220906 \\
      S\&P   & 0.158347 & 0.552242 & 0.675382 & 0.197728 \\
      \bottomrule
      \end{tabular}%
    \label{tab:sec_diffsharpe_results}%
\end{table}%

\begin{table}[htbp]
    \centering
    \caption{Strategies with SEC Filings (Profit Reward)}
      \begin{tabular}{lcccc}
      \toprule
            & Net Profit & Sharpe Ratio & Sortino Ratio & Max Drawdown \\
      \midrule
      SEC CNN & 0.162534 & 0.532298 & 0.649254 & 0.195766 \\
      SEC RNN & 0.154506 & 0.513757 & 0.626332 & 0.187671 \\
      SEC MLP & 0.073242 & 0.266018 & 0.329634 & 0.208675 \\
      S\&P   & 0.158347 & 0.552242 & 0.675382 & 0.197728 \\
      \bottomrule
      \end{tabular}%
    \label{tab:sec_profit_results}%
\end{table}%

SEC data is regularly available for all companies in our universe and is of high quality. As a result, we see significant performance improvement. In Table \ref{tab:sec_diffsharpe_results}, results are strong across the board using the Differential Sharpe reward, but none quite match the S\&P index. Differential Sharpe is a difficult reward, so it imposes a penalty on optimal performance; however, it is difficult for a model to overfit, so none of the returns are abysmal. In Table \ref{tab:sec_profit_results}, we see impressive results from the CNN and RNN EIIE policies, but bad performance from the MLP policy. This is likely because the overly complex MLP policy overfits and suffers. The CNN and RNN policies using the Differential Sharpe ratio are among the strongest contenders in our experiments.

\subsection{RL Portfolios: Price + SEC + News Data}

Unfortunately, the news data is much less regular than the SEC data and is not available consistently for all stocks in our universe. While it does provide some improvement over only using historical price data as in Section \ref{sec:hist_price_results}, it is not as significant as with SEC data.

\begin{figure}[!htb]
    \centering
    \includegraphics[width=0.8\textwidth]{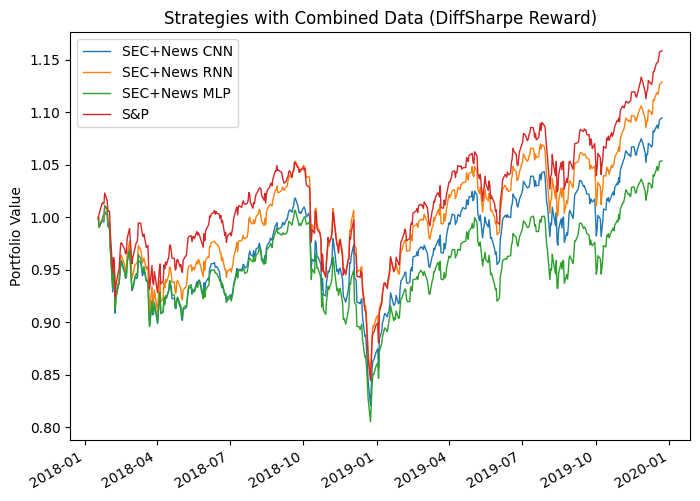}
    \caption{Performance of strategies using price, SEC, and news data with Differential Sharpe reward}
    \label{fig:comb_diffsharpe_results}
\end{figure}

\begin{figure}[!htb]
    \centering
    \includegraphics[width=0.8\textwidth]{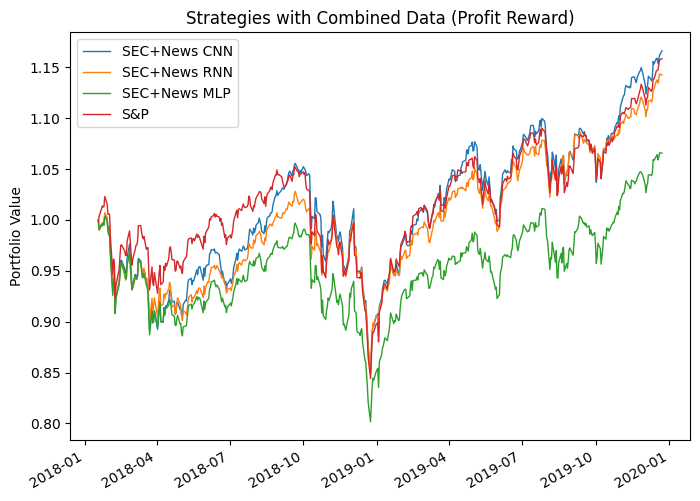}
    \caption{Performance of strategies using all price, SEC, and news data with Profit reward}
    \label{fig:comb_profit_results}
\end{figure}

However, combining the price, SEC, and news sentiment data provides strong results. The performance of models trained on this dataset with the Differential Sharpe Ratio reward can be viewed in Table \ref{tab:combined_sharpe} and Figure \ref{fig:comb_diffsharpe_results}, while results for the Profit reward are found in Table \ref{tab:combined_profit} and Figure \ref{fig:comb_profit_results}.

\begin{table}[!htb]
    \centering
    \caption{Strategies with Combined Data (DiffSharpe Reward)}
      \begin{tabular}{lcccc}
      \toprule
            & Net Profit & Sharpe Ratio & Sortino Ratio & Max Drawdown \\
      \midrule
      SEC+News CNN & 0.094546 & 0.342485 & 0.412857 & 0.193996 \\
      SEC+News RNN & 0.128636 & 0.444354 & 0.536047 & 0.195274 \\
      SEC+News MLP & 0.053514 & 0.205204 & 0.248546 & 0.200069 \\
      S\&P   & 0.158347 & 0.552242 & 0.675382 & 0.197728 \\
      \bottomrule
      \end{tabular}%
    \label{tab:combined_sharpe}%
\end{table}%

\begin{table}[!htb]
    \centering
    \caption{Strategies with Combined Data (DiffSharpe Reward)}
      \begin{tabular}{lcccc}
      \toprule
            & Net Profit & Sharpe Ratio & Sortino Ratio & Max Drawdown \\
      \midrule
      SEC+News CNN & 0.166081 & 0.553644 & 0.664706 & 0.195472 \\
      SEC+News RNN & 0.142287 & 0.519258 & 0.636281 & 0.170134 \\
      SEC+News MLP & 0.065676 & 0.248586 & 0.304594 & 0.195750 \\
      S\&P   & 0.158347 & 0.552242 & 0.675382 & 0.197728 \\
      \bottomrule
      \end{tabular}%
    \label{tab:combined_profit}%
\end{table}%

As seen in Table \ref{tab:combined_sharpe}, the policies trained for Differential Sharpe Ratio do not fare as well because the irregularity of the news makes a bigger impact on the harsher reward. Thus, we end up seeing similar results to Section \ref{sec:hist_price_results}, with the CNN and RNN policies performing better than the MLP but still mediocre.

However, for the profit reward, we do see a notable increase in Table \ref{tab:combined_profit}. Indeed, our best-performing model across all three sets of experiments is the CNN EIIE model on the combined dataset using Profit reward. Again, the MLP dramatically overfits.

\subsection{Comparison}
\begin{figure}[!htb]
    \centering
    \includegraphics[width=0.8\textwidth]{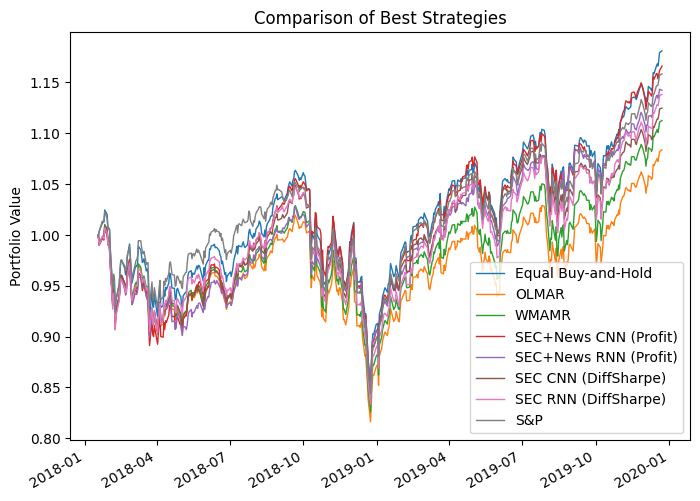}
    \caption{Comparison of Best Strategies}
    \label{fig:best_compare}
\end{figure}

\begin{figure}[!htb]
    \centering
    \includegraphics[width=0.8\textwidth]{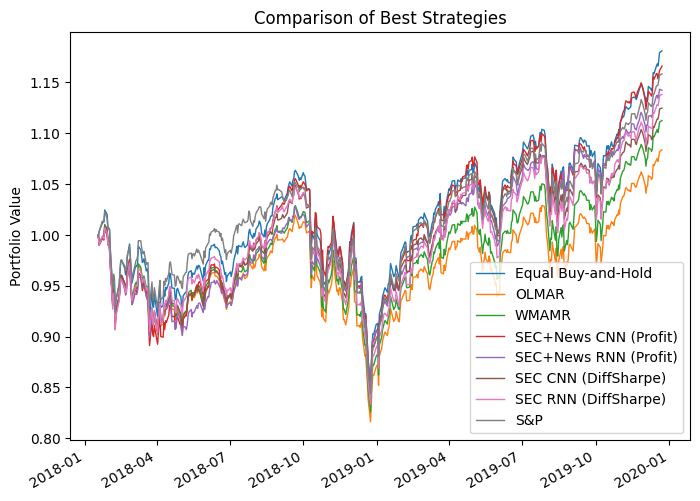}
    \caption{Comparison of Best Strategies (with rescaled y-axis)}
    \label{fig:best_compare_scaled}
\end{figure}

Figures \ref{fig:best_compare} and \ref{fig:best_compare_scaled} and Table \ref{tab:benchmark_comparison} compare the performance of the best models we trained against select benchmarks from Section \ref{sec:benchmark_section}.

\begin{table}[htb]
    \centering
    \begin{tabular}{lcccc}
        \toprule
        & Net Profit & Sharpe Ratio & Sortino Ratio & Max Drawdown \\
        \midrule
        Equal Buy-and-Hold & 0.181138 & 0.595225 & 0.716195 & 0.198039 \\
        OLMAR & 0.083634 & 0.300906 & 0.364244 & 0.199240 \\
        WMAMR & 0.112565 & 0.392310 & 0.473818 & 0.197294 \\
        SEC+News CNN (Profit) & 0.166081 & 0.553644 & 0.664706 & 0.195472 \\
        SEC+News RNN (Profit) & 0.142287 & 0.519258 & 0.636281 & 0.170134 \\
        SEC CNN (DiffSharpe) & 0.124870 & 0.439863 & 0.529951 & 0.195347 \\
        SEC RNN (DiffSharpe) & 0.138309 & 0.464039 & 0.568351 & 0.204159 \\
        S\&P & 0.158347 & 0.552242 & 0.675382 & 0.197728 \\
        \bottomrule
    \end{tabular}
    \caption{Comparison of Best Strategies against Benchmarks}
    \label{tab:benchmark_comparison}
\end{table}

Excluding Equal Buy-and-Hold, our SEC+News CNN EIIE policy with Profit reward has the highest net profit, Sharpe ratio, and Sortino ratio. Additionally, the SEC+News RNN policy with the profit reward has the lowest max drawdown of all strategies. All of our trained strategies outperform the OLMAR and WMAMR trading benchmarks.

\section{Conclusion}

Our principal observations indicate a disparity in learning complexity between the profit reward function and the differential Sharpe ratio for our agent, resulting in consistently superior portfolio performance when optimizing with the former. 
Notably, agents trained on CNN EIIE and RNN EIIE, exhibit enhanced performance under the profit reward, while the MLP policy network seems to overfit significantly. This is because compared to the MLP, both CNNs and RNNs have fewer weights and biases to learn.

Furthermore, our analysis underscores the challenge of integrating news data, revealing a diminished capacity of the model to learn the differential Sharpe ratio reward. We attribute this difficulty to the sparse and inconsistent nature of the dataset. 

Upon integrating SEC filings data, notable performance enhancements are observed across both reward functions compared to baseline models. The regularity and consistency of SEC data across all tickers facilitate improved learning for our policy algorithms. News data also provides benefits when used in combination with SEC data. However, the irregularities in the data have a significant impact when using a difficult reward, such as the Differential Sharpe Ratio. Optimal performance is achieved when simultaneously integrating news and SEC data into the agent's environment. This outcome underscores the untapped potential of research avenues exploring comprehensive datasets capturing nuanced company sentiment. 
With a larger and more consistent news headline dataset, we believe that we could create better-performing agents. However, the challenges in using alternative data can be somewhat reduced by choosing an appropriate reward and a policy that is sufficiently resistant to overfitting.

Furthermore, there are other routes to improve returns beyond simply improving the quality of our data. For instance, we could test different feature extractors and apply different regularization techniques. In addition, different sentiment embedding functions could potentially be more accurate and/or usable by our agents.

\newpage

\bibliographystyle{unsrt}  
\bibliography{references}


\newpage

\section*{Appendix A}

\subsection*{Reinforcement Learning Overview}
\label{sec:rl_overview}

The reader may not be readily familiar with reinforcement learning (RL).
Thus, we here provide a brief overview of the terminology, basic definition, essential results, and common algorithms in RL theory.

\subsubsection*{Markov Decision Process}

A Markov Decision Process (MDP) problem is a framework for modeling sequential decision-making by an agent in an environment.
A problem is formally defined as a $4$-tuple $(S, A, T, R)$.
\begin{itemize}
  \item $S$ is the state space, which is the set of all possible states of the environment.
  \item $A$ is the action space, which contains all possible actions that the agent can take (across all possible states).
  \item $T: S \times A \times S \to [0, 1]$ is the transition function in a stochastic environment. When the environment is in state $s$ and
    the agent takes action $a$, then $T(s, a, s')$ is the probability that the environment transitions to state $s'$ as a result. (In a deterministic environment,
    this function may not be necessary, as there may be only one possible state due to the taken action.)
  \item $R: S \times A \times S \to \R$ is the reward function. When the environment changes state from $s$ to $s'$ due to action $a$, the agent receives reward $R(s, a, s')$.
\end{itemize}
The environment is Markov, which means that the distribution of the next state $s'$ conditioned on the current state $s$ and action $a$ is independent of the time step.

A policy is a function $\pi: S \to A$ that dictates actions to take at a given state.
A solution to an MDP problem is an optimal policy $\pi^*$ that maximizes the agent's utility, however, that is defined.

\subsubsection*{RL Terminology}

In RL, the agent's utility is generally defined as the total expected discounted reward.
Let $\gamma \in [0, 1]$ be a constant discount factor. The utility from a sequence of reward $\{r_t\}_{t=0}^\infty$ is
thus commonly defined as $U([r_0, r_1, \ldots]) = \sum_{t=0}^\infty \gamma^t r_t \leq \frac{\sup_t r_t}{1-\gamma}$.
The benefit of this formulation is that (1) utility is bounded if the rewards are bounded, and (2) there is a balance
between small immediate rewards and large long-term rewards. (The use of the discount factor depends on the actual reward function.
For custom reward functions, it may not be necessary or even desirable; we include it because it is common in RL literature.)

Given a policy $\pi: S \to A$, we define the value function $V^\pi: S \to \R$ and the $Q$-function $Q^\pi: S \times A \to \R$ as
\begin{align*}
  V^\pi(s) = \E_\pi\left[ \sum_{t=0}^\infty \gamma^t R_t \:\Big|\: s_0 = s  \right] &&
  Q^\pi(s, a) = \E_\pi\left[  \sum_{t=0}^\infty \gamma^t R_t \:\Big|\: s_0 = s, a_0 = a \right]
\end{align*}
$V^\pi(s)$ is the expected utility from starting at $s$ and following policy $\pi$, and $Q^\pi(s, a)$ is the expected utility
from starting at $s$, taking action $a$, and then following policy $\pi$ thereafter.
The goal of RL is to find the optimal policy $\pi^*$, from which we have the optimal value function $V^*$ and optimal $Q$-function $Q^*$.
These optimal values can further be defined as follows:
\begin{align*}
  Q^*(s, a) &= \E_{s'}[R(s, a, s') + \gamma V^*(s')] = \sum_{s'} T(s, a, s')[R(s, a, s') + \gamma V^*(s')] \\
  V^*(s) &= \max_a Q^*(s, a) = \max_a \sum_{s'} T(s, a, s')[R(s, a, s') + \gamma V^*(s')]
\end{align*}
The second equation is known as the Bellman equation.

There are two main branches of RL: model-based and model-free.
In model-based RL, the agent attempts to build a model of the environment transition function $T$ and reward function $R$.
Based on this model, it then attempts to directly maximize the total expected reward.
In model-free RL, the agent does not attempt to model the environment but instead attempts to learn either the value function or $Q$-function.
Once it has one of these, it can derive an optimal policy from it:
\begin{align*}
  \pi^*(s) = \arg\max_a Q^*(s, a) = \arg\max_a \sum_{s'} T(s, a, s')\left[ R(s, a) + \gamma V^*(s')\right]
\end{align*}
We proceed with model-free RL in this project.

\newpage

\section*{Appendix B}
\label{AppendixB}

\subsection*{OLMAR Benchmark Strategy}
\label{olmar_section}
Online Portfolio Selection with Moving Average Reversion (OLMAR), introduced in \cite{olmar_paper}, is used by both \cite{rl_augmented_states} and \cite{drl_framework} as a benchmark starting strategy.
Our codebase includes both custom and library implementations of OLMAR; we provide a brief description of the method here.

Consider a universe of $m$ tradable assets (and no risk-free asset).
Let $$\Delta_m = \left\{b_t \in \R^m \:|\: b_{t,i} \geq 0 \:\forall i, \sum_{i=1}^m b_{t,i} = 1\right\}$$ be the space of possible portfolios, which is a simplex.
Let $p_t \in \R^{m}$ be the price at time $t$, and and let $x_t \in \R^m$ be the price-relative vector for time $t$, computed as
$x_{t,i} = p_{t,i} / p_{t-1,i}$ for all $i$; that is, $x_t$ is the element-wise division of $p_t$ by $p_{t-1}$.
The goal of the algorithm is to produce a good portfolio $b_{t+1}$ given $p_t, p_{t-1}, ... p_{t-w+1}$ (that is, historical stock prices over a lookback of period $w$).

If we believe that the price of an asset is mean-reverting, then a good prediction for the next price-relative vector $\tilde x_{t+1}$ is
\begin{align*}
  \tilde x_{t+1} = \frac{\mathop{\mathrm{MA}}_t(x)}{p_t} = \frac{1}{w}\left(\frac{p_t}{p_t} + \frac{p_{t-1}}{p_t} + \cdots + \frac{p_{t-w+1}}{p_t}\right)
\end{align*}
To obtain a good return over the next period, we want $b_{t+1} \cdot \tilde x_{t+1}$ to be high.
However, to keep transaction costs down, we do not want to be too far away from the previous portfolio $b_t$.
Therefore, we formula the optimization problem:
\begin{align*}
  b_{t+1} = \mathop{\mathrm{arg\:min}}_{b \in \Delta_m} \frac{1}{2}\|b - b_t\| \:\:\text{such that}\:\: b \cdot \tilde{x}_{t+1} \geq \epsilon
\end{align*}
for some positive threshold value $\epsilon$. This optimization problem can be solved numerically using standard-constrained solvers.

Alternatively, the authors present the following solution: if we ignore the non-negativity constraint, then the solution is
\begin{align*}
  b_{t+1} = b_t + \lambda_{t+1}(\tilde x_{t+1} - \bar x_{t+1} \mathbb{1}) && \bar x_{t+1} = \frac{\mathbb{1} \cdot \tilde x_{t+1}}{m} && \lambda_{t+1} = \max\left\{0, \frac{\epsilon - b_t \cdot \tilde x_{t+1}}{\|\tilde x_{t+1} - \bar x_{t+1} \mathbb{1}\|^2}\right\}
\end{align*}
To enforce the non-negativity constraint, we can project this solution back into the simplex $\Delta_m$.

\subsection*{WMAMR Benchmark Strategy}
\label{wmamr_section}
Weighted Moving Average Mean Reversion, introduced in \cite{wmamr_paper}, is another trading strategy benchmark used by \cite{rl_augmented_states}.
It is a relatively simple modification to OLMAR, so we will not restate the content of Section \ref{olmar_section}.

The definitions of all terms remain the same. Define the $\epsilon$-insensitive loss function
\begin{align*}
  l_{1,\epsilon}(b, \tilde x_{t+1}) = \begin{cases}
    0 & b \cdot \tilde x_{t+1} \leq \epsilon \\
    b \cdot \tilde x_{t+1} - \epsilon & \text{otherwise}
  \end{cases}
\end{align*}
The authors formulate the optimization problem as
\begin{align*}
  b_{t+1} = \mathop{\mathrm{arg\:min}}_{b \in \Delta_m} \frac{1}{2}\|b - b_t\|^2 \:\:\text{such that}\:\: l_{1,\epsilon}(b, \tilde x_{t+1}) = 0
\end{align*}
As with OLMAR, this optimization problem can be solved numerically.
If we ignore the non-negativity constraint, then an analytic solution is
\begin{align*}
  b_{t+1} = b_t - \tau_t(\tilde x_{t+1} - \bar x_{t+1} \cdot \mathbb{1}) && \tau_t = \max\left\{0, \frac{l_{1,\epsilon}}{\|\tilde x_{t+1} - \bar x_{t+1}\|^2}\right\}
\end{align*}
To recover the non-negativity constraint, we project this solution back into the simplex $\Delta_m$.

\newpage

\section*{Image Gallery}
\label{chap:gallery}

\end{document}